\documentclass[twocolumn]{IEEEtran}
\usepackage{epsfig,amsmath,amstext,amsfonts,amssymb,cite,cases,multirow}
\usepackage{algorithm} 
\usepackage{algorithmic} 
\usepackage{mathrsfs}
\usepackage{color}

\usepackage{amsmath}
\usepackage{graphicx}
\usepackage{subfigure}
\usepackage{bm}
\usepackage{subeqnarray}
\usepackage{cases}
\usepackage{diagbox}
\usepackage{makecell}

\begin{document}
	
	\title{Peak-to-Average Power Ratio Analysis for OFDM-Based Mixed-Numerology Transmissions}
	\author{
		Xiaoran~Liu, ~
		Lei~Zhang,~\IEEEmembership{Senior~Member,~IEEE,}
		Jun~Xiong,~\IEEEmembership{Member,~IEEE,}
		Xiaoying~Zhang,
		Li~Zhou 
		and~Jibo~Wei,~\IEEEmembership{Member,~IEEE}
		
		
	}
	
	\maketitle
	
	\begin{abstract}
		In this paper, the probability distribution of the peak to average power ratio (PAPR) is analyzed for the mixed numerologies transmission based on orthogonal frequency division multiplexing (OFDM).
			State of the art theoretical analysis implicitly assumes continuous and symmetric frequency spectrum of OFDM signals.
			Thus, it is difficult to be applied to the mixed-numerology system due to its complication.
			By comprehensively considering system parameters, including numerology, bandwidth and power level of each subband, we propose a generic analytical distribution function of PAPR for continuous-time signals based on level-crossing theory.
		The proposed approach can be applied to both conventional single numerology and mixed-numerology systems.
		In addition, it also ensures the validity for the noncontinuous-OFDM (NC-OFDM).
		Given the derived distribution expression, we further investigate the effect of power allocation between different numerologies on PAPR.
		Simulations are presented and show the good match of the proposed theoretical results.
	\end{abstract}
	
	
	\section{Introduction}\label{sec1}
	Due to its advantages of high spectral efficiency and robustness against multipath fading, 
	orthogonal frequency division multiplexing (OFDM) has been widely used in many wireless communication standards, such as LTE/LTE-A, IEEE 802.11 and forthcoming fifth generation New Radio (5G NR)\cite{2017-3GPPTechnical}.
	The prime drawback of OFDM is its high peak-to-average power ratio (PAPR), because of the inherent summation of multiple parallel data flows transmitted on different subcarriers.
	Hence, the dynamic range of power amplifier (PA) is required to be large enough to avoid the nonlinear distortion that may severely impair the OFDM signals and degrade receiver performance\cite{2013ISPM-WunderPAPR}.
	
	Recently, the mixed-numerology transmission has been proposed in 5G NR\cite{2017-3GPPTechnical}. 
	Compared with the conventional single numerology system that only uses one unified set of parameters for all subcarriers (such as LTE/LTE-A and 802.11), 
	the mixed-numerology system adopts different waveform parameters, such as cycle prefix and subcarrier spacing,
	to support different services and use cases.
	For example, Vehicle to everything (V2X) communications may suffer from serious Doppler effect and has stringent latency requirement.  
	{\color{black}The large subcarrier spacing which is more robust to frequency spread and has smaller symbol duration is thus preferred.
		Machine-type communications, on the other hand, require smaller subcarrier spacing to support very large number of devices within limited bandwidth\cite{2016ICM-Waveform}.}
	With configuring multiple numerologies in time-frequency domains, mixed-numerology system enables the adaptive selection of numerology\cite{2018IA-YazarFlexibility} and the flexible scheduling\cite{Gonzalez}, based on the channel conditions and the quality of service (QoS) demands.
	However, as the OFDM technique is reserved as the basic waveform\cite{2017-3GPPTechnical}, the PAPR problem still exists and becomes more complex and challenging  \cite{2016ICM-Waveform,2016ICM-ZhangWaveform,2017ICM-Lien5G}.
	
	For the convenience of the design and evaluation of system parameter and PAPR reduction schemes, it is of great importance to identify the PAPR distribution of OFDM signals.
	For example, the output back-off of PA can be determined according to the PAPR distribution\cite{2013ICST-RahmatallahPeak}.
	It also contributes toward the derivation of the approximate signal error rate as well as achievable information rate \cite{2017ITWC-Modeling,2000ITIT-Frieseachievable}.
	Furthermore, 
	the understanding of the PAPR property is critical in PAPR reduction schemes, such as block partial transmit sequence (PTS)\cite{2001ICL-TellamburaImproved} and selected mapping method (SLM)\cite{2010IJSTSP-low},
	in which the corresponding parameters can be properly designed based on PAPR.
	Therefore, analytical expression of PAPR distribution is essential to facilitate the system design.
	
	In general, PAPR distribution can be either bounded and theoretically derived.
	{\color{black}The former approach is to seek lower or upper bounds of the PAPR distribution.
		However, these bounds may be far deviated from the practical scenario.
		As a result, the derivation of PAPR distribution is usually preferred to take into account of statistical characterization.}
	By assuming the transmitted OFDM signal as a complex Gaussian random process,
	several theoretical approximate expressions have been derived in \cite{2001ITC-Ochiaidistribution,2002-modern,2008ITWC-Derivation}, which provide good approximation to the simulation results.
	Meanwhile, the assumption that the transmitted OFDM signal asymptotically follows Gaussian random process is rigorously proved in \cite{2010ITIT-Convergence}.
	More recently, the PAPR analysis with low subcarriers number has been presented in \cite{2018ITC-PAPR}, which can be applied for the narrowband Internet of things (NB-IoT) system.
	
	Unfortunately, the existing PAPR analysis of single numerology cannot be directly applied for mixed-numerology system due to the following reasons.
	First, the PAPR of mixed-numerology signals is measured for the sum of OFDM signals from each subband with different numerologies.
	The PAPR of such composite signals is therefore affected by more complicated system parameters than the conventional OFDM signals.
	Second, the results of previous works \cite{2001ITC-Ochiaidistribution,2002-modern,2008ITWC-Derivation,2018ITC-PAPR} are based on the case that the frequency spectrum of OFDM signal is symmetric and continuous, which in general are not true in mixed-numerology system.
	{\color{black}
		In addition, there are very few researches on the PAPR distribution analysis of noncontinuous-OFDM (NC-OFDM)\cite{2017ITVT-NiError,2015ITVT-NiJoint,2014ICM-JiangEnergy,2007-Peak,2017ToETT-PAPR},
	}
	which only includes the simulation discussion \cite{2007-Peak} and bound derivation \cite{2017ToETT-PAPR}.
	Third, the theoretical analysis of PAPR distribution with small number of subcarriers may become inaccurate when the oversampled signals are considered\cite{2018ITC-PAPR}.
	Therefore, we need to reconsider the PAPR analysis for the mixed-numerology system.
	
	The main contributions of this paper are summarized as follows.
	
	{\color{black}
		\begin{itemize}
			\item 
			We first define the PAPR for OFDM-based mixed-numerology system. 
			Considering the system parameters, including numerology, bandwidth and power level, we derive an expression of PAPR distribution for continuous-time signals with the help of the level-crossing theory of random processes.
			The proposed expression can be regarded as a generalization of the existing single numerology PAPR analysis\cite{2001ITC-Ochiaidistribution}.
			Additionally, 
			our method is also applicable to NC-OFDM system.
			\item 
			We also investigate the impact of power allocation on PAPR performance,
			where the power allocation among different subbands is formulated as an optimization problem.
			In particular, when two subbands are adopted, we provide a closed-form solution.
			The derived result shows that PAPR is largely determined by the bandwidth regardless of the selection of numerology.
		\end{itemize}
	}
	
	The remainder of the paper is organized as follows. 
	Section II presents a general system model of OFDM-based mixed-numerology transmission and identifies the PAPR problem. 
	Section III proposes a general analytical expression of PAPR distribution for mixed-numerology system.
	The effect of power allocation on PAPR is also included in this section.
	The proposed theoretical results are validated through simulations in Section IV. 
	Finally, several concluding remarks are made in Section V. 
	Main proof is given in Appendix to maintain the flow of the paper.
	
	\begin{figure}[t]
		\centering
		\includegraphics[width=90mm]{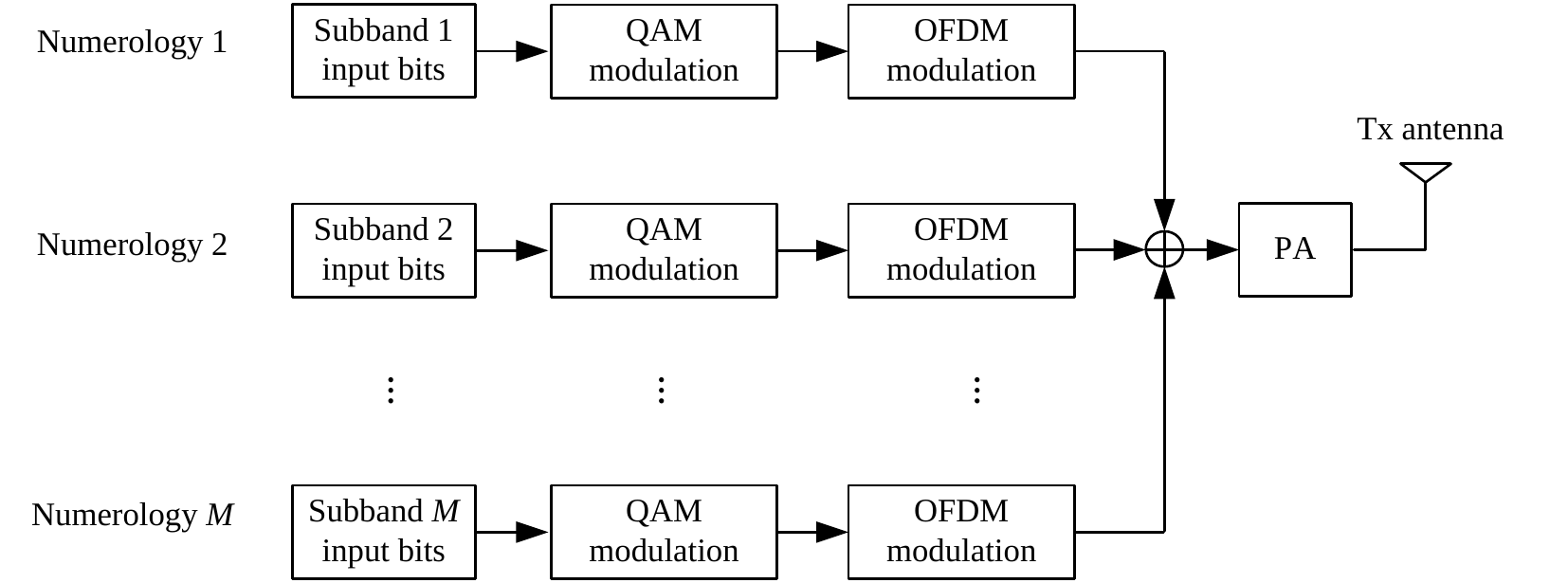}
		\caption{Block diagram of OFDM-based mixed-numerology transmitter.}
		\label{model}
	\end{figure}
	
	{\it{Notations:}}
	Boldface lowercase and uppercase letters denote column vectors and matrices, respectively.
	Superscripts $ \{\cdot\}^T $ stand for transpose operation.
	$ {\bf{1}}_m $ refers to $ m $-element all-ones column vector.
	The notations $ E[\cdot] $, $ \operatorname{rank}(\cdot) $, $ \operatorname{Pr}(\cdot) $ represent the expectation, rank and probability, respectively.
	$ {\mathcal{Z}} $ and $ {\mathcal{Z}}^+ $ denotes the set for non-negative and positive integers, respectively.
	$ {\mathcal{CN}}(0,\sigma^2) $ denotes the complex Gaussian distribution with zero mean and variance $ \sigma^2 $.

	\section{System Model}\label{sec2}

	We consider an OFDM-based mixed-numerology transmitter as shown in Fig. \ref{model}.
	The baseband processing diagram illustrates the coexistence of $ M $ numerologies.
	The system bandwidth is split into several subbands in which diverse numerologies (with different subcarrier spacings) may be applied to support dedicated services.
	To simplify the system model, we assume that each numerology is adopted only in one subband.
	\footnote{{\color{black}When different numerologies are adopted in one subband, it can be seen as the combination of multiple subbands with corresponding numerologies.
			Hence, this simplification is no loss of generality.}}
	Then, all subband signals are added together before being sent to PA for transmission.
	
	For the $ i $-th subband with $ N_i $ continuous subcarriers using subcarrier spacing $ f_i $, the $ u $-th OFDM signal is expressed as 
	\begin{equation}\label{OFDMsymbol}
	\begin{split}
	s_{i}^u(t) = \sqrt{{\eta_i} \over {N_i}}\sum_{k = 0}^{{N_i } - 1}A_ke^{j2\pi (k f_i+\delta_i){\color{black}(t - T_{C\!P,i}-(u-1)T_i)}}, \\
	\quad  (u-1)T_i \leq t < uT_i
	\end{split}
	\end{equation}
	where $ A_k $ is the complex information symbol carried by the $ k $-th subcarrier and 
	is assumed to be statistically independent, identically distributed (i.i.d.) random variables with zero mean and unit variance, i.e., $ E[|A_k|^2]=1 $.
	The average power of the $ i $-th subband is $ \eta_i $.
	{\color{black}The symbol duration is $ T_i = T_{C\!P,i} + T_{sys,i} $ where $ T_{C\!P,i} $ is the cyclic prefix (CP) length and $ T_{sys,i}=1/f_i $.}
	$ \delta_i $ is the frequency of the carrier with the lowest frequency.

\begin{figure}[t]
		\centering
		\includegraphics[width=88mm]{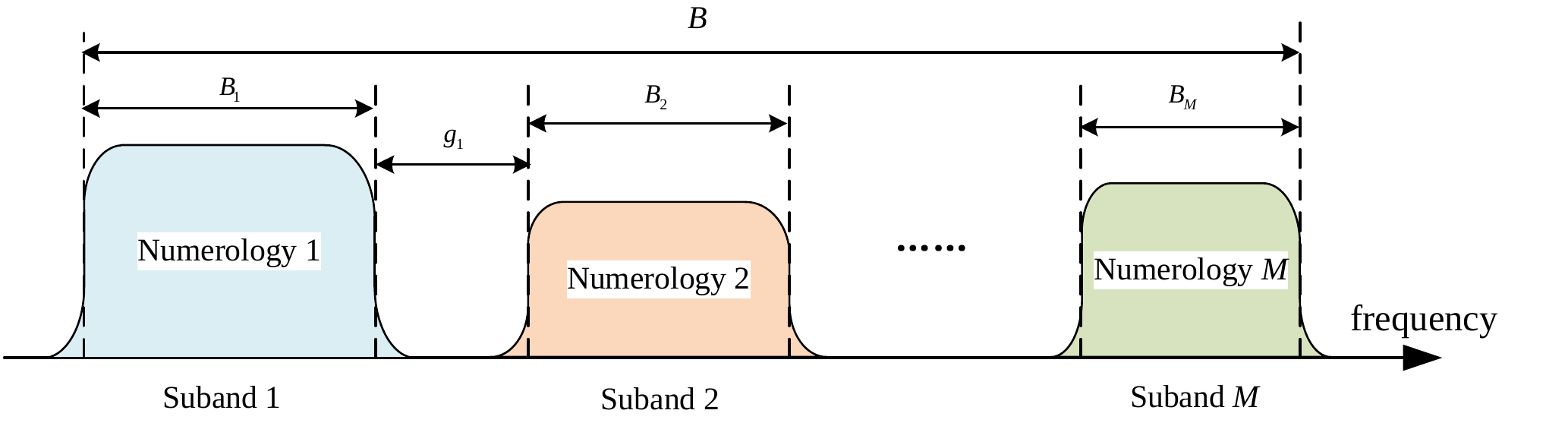}
		\caption{Frequency-domain representation of mixed-numerology signal.}
		\label{psd}
	\end{figure}

	Fig. \ref{psd} indicates the frequency spectrum of $ M $ subbands in the system bandwidth, each of which has the bandwidth $ B_i $.  
	Because the power spectrum density (PSD) of OFDM signal is approximately rectangular, the subband bandwidth can be determined by the number of subcarriers, i.e., $ B_i \approx N_i f_i  $\cite{2001ITC-Ochiaidistribution,2008ITWC-Derivation}.
	Suppose that subbands are not overlapped and the guard interval should be reserved from adjacent subband to avoid the interference.
	The guard interval between $ i $-th and $ i\!+\!1 $-th subbands is denoted as {\color{black}$ g_i $}.
	Then, the frequency shift of the $ i $-th subband is
	$ \delta_i=\sum_{v=1}^{i-1}(B_v+g_v) $
	and the system bandwidth $ B $ can be roughly calculated as
	\begin{equation}\label{}
	B = \sum_{v=1}^{M-1}(B_v+g_v) + B_M.
	\end{equation}
	
	We now consider a generalized synchronized system {\color{black}in which all subbands share an integral least common multiple (LCM) duration $ T_{0} $ and starting time instance} \cite{2017ICM-ZhangMulti,2017IToWC-ZhangSubband}:
	\begin{equation}\label{period}
	T_{0} = n_1 T_1 = n_2T_2 = ...=n_MT_M, 
	\end{equation}
	where $ n_i  \in {\mathcal{Z}}^+$ for $ i=1,2,...,M $.
	{\color{black}Thus, one LCM symbol can be considered as a closed space in which only limited symbols need to be processed, and each LCM symbol has the same overall composition which can be easily handled for performance analysis.}
	
	\begin{figure}[t]
		\centering
		\includegraphics[width=38mm]{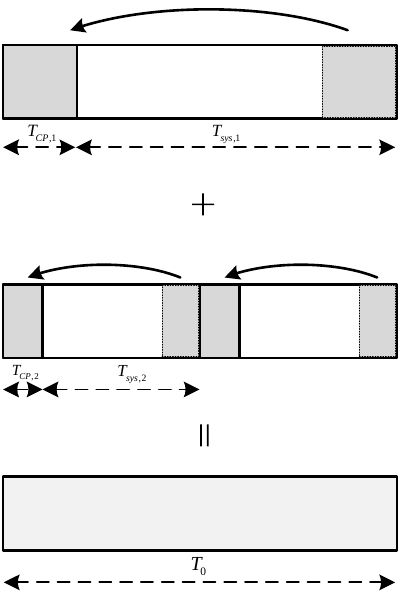}
		\caption{{\color{black}Example of composite signal with two numerologies.}}
		\label{symbol}
	\end{figure}
	A commonly accepted family of numerologies are defined as follows \cite{2017-3GPPTechnical,2017IJoSAiC-Guan5G,2016ICM-Waveform}:
	\begin{equation}\label{scs}
	f_{i} = 2^{L_i} f_{1}, \, T_{C\!P,i}=T_{C\!P,1}/2^{L_i},
	\end{equation}
	where $ L_i \in {\mathcal{Z}} $ and {\color{black}$ f_{1}, T_{C\!P,1} $ are the subcarrier spacing and CP length of the base numerology, respectively.}
	Without loss of generality, we assume $ f_{1} $ as the minimum subcarrier spacing among $ M $ numerologies.
	Eq. (\ref{scs}) shows that any subcarrier spacing is an integer
	divisible by all smaller subcarrier spacing,
	which implies that different OFDM numerologies can be implemented by using different-scaled IFFT under the same sampling clock rate\cite{2017IJoSAiC-Guan5G,2017ICM-ZhangMulti}.
	
	In generalized synchronized system, the LCM duration is therefore seen as $ T_{0} = T_1 $ and  
	\begin{equation}\label{}
	n_1=1,\quad n_i = 2^{L_i}, i = 2,3,...,M.
	\end{equation}
	By this mean, a mixed-numerology transmitted signal, 
	defined over the time interval $ t\in[(u-1)T_0,uT_0] $,
	is a composite of multiple OFDM signals from $ M $ subbands, each of which have $ n_i $ OFDM signals for $ i=1,2,...,M $, i.e.,
	\begin{equation}\label{z}
	z^u\left( t \right) = \sum_{i=1}^{M} \sum_{v=0}^{n_i-1}{s_{i}^{u_i}}\left( t \right), (u-1)T_0 \leq t < uT_0
	\end{equation}
	where $ u_i=n_iu+v $.
	{\color{black}For example, when two subbands with $ 2f_1=f_2 $ are considered, the composition of one LCM symbol can be shown in Fig. \ref{symbol}.}
	
	For the convenience of the following illustration, we omit the superscript $ u $.
	The total average power of composite signal is
	\begin{equation}\label{}
	P_{{\rm av}} = E[|z(t)|^2] = \sum_{i=1}^{M}{\eta_i}.
	\end{equation}

	According to the definition of PAPR which is the ratio of the peak power of the signal to its average power, we can define the PAPR of composite signal $ z(t) $ as
	\begin{equation}\label{PAPR}
	\gamma = { \dfrac{\max_{t \in [0, T_0)}  \left\vert z(t)\right\vert^{2}}{P_{{\rm av}}} }.
	\end{equation}
	Without loss of generality, $ P_{{\rm av}} $ is normalized to be one in the rest of this paper.
	
	{\it{Remark 1:}}
	From (\ref{PAPR}), it is noteworthy that the PAPR of the composite signal $ z(t) $ is evaluated for the summation of all the subband signals.
	Therefore, in order to simplify the system design, it is necessary to understand the properties of the PAPR in mixed-numerology system.
	For example, the system parameters, such as the bandwidth division, numerology selection and power allocation, can be taken into consideration to evaluate the PAPR performance.
	In addition, we can determine the reasonable back-off point of power according to the saturation of PA to effectively reduce the nonlinear distortion.
	

	\section{Analysis of PAPR Distribution for Mixed-Numerology System}\label{sec3}
	\subsection{{\color{black}Related Works}}
	{\color{black}
		The PAPR distribution of the conventional OFDM signals has been widely studied in the previous works.
		When the OFDM signals are sampled at the Nyquist rate, the CCDF of PAPR has been given in \cite{1998Nee}
		\begin{equation}\label{ccdf_nyq}
		\begin{split}
		{\rm{CCDF}}(\gamma)= 1-(1-e^{-\gamma})^{N},
		\end{split}
		\end{equation}
		where $ N $ is the number of subcarriers.
		The CCDF obtained in Nyquist sampling rate generally underestimates the PAPR distribution of the continuous-time signals.
		For more practical scenarios, an empirical approximation is then proposed \cite{1998Nee} 
		\begin{equation}\label{emp}
		{\rm{CCDF}}(\gamma)=1-(1-e^{-\gamma})^{2.8 N},
		\end{equation}
		where $ 2.8 $ is an empirical parameter.
		However, the empirical expression (\ref{emp}) still leads to discrepancies with simulation results and the researchers are more interested with new methods that have theoretical justification.

		In \cite{2001ITC-Ochiaidistribution}, the analytical expression of PAPR distribution is developed for the continuous-time OFDM signals.
		Based on the level crossing theory\cite{1944BSTJ-RiceMathematical}, the derived expression is written as
		\begin{equation}\label{R1}
		{\rm{CCDF}}(\gamma) \approx 1 - \exp \left(- N e^{-\gamma} \sqrt{\dfrac{\pi}{3}\gamma}\right),
		\end{equation}
		The theoretical result is able to achieve good accuracy when the number of subcarriers is relatively large.
		However, the derivation in \cite{2001ITC-Ochiaidistribution} implicitly assumes the OFDM signal continuous and symmetric in frequency domain so that the obtained result cannot be applied to more general cases.
		Furthermore, the power allocation problem is not considered, which have a potential impact on the PAPR distribution.
		
		In \cite{2002-modern}, based on the extreme value theory, a simple approximate CCDF of PAPR is derived as
		\begin{equation}\label{R2}
		{\rm{CCDF}}(\gamma) \approx 1 - \exp \left(- N e^{-\gamma} \sqrt{\dfrac{\pi}{3}\log{N}}\right) .
		\end{equation}
		This work has been further developed for considering the power distribution on different subcarriers in \cite{2008ITWC-Derivation}.
		It gives the approximation as
		\begin{equation}\label{R3}
		{\rm{CCDF}}(\gamma) \approx 1 - \exp \left(- 2 e^{-\gamma} \sqrt{\dfrac{\pi \gamma}{NP_{av}}\sum_{k=-N/2}^{N/2}k^2\varepsilon_k}\right),
		\end{equation}
		where $ \varepsilon_k $ is the power allocated on the $ k $-th subcarrier.
		The basic idea behind \cite{2002-modern,2008ITWC-Derivation} requires that the frequency spectrum of the OFDM signal is symmetric so that 
		the envelope of OFDM signal is a special Chi-squared-2 process in which the autocorrelation functions can be expressed by the expansion $r\left( t \right) = 1 - \left( { - r''\left( 0 \right)} \right)\frac{{{t^2}}}{2} + o\left( {{t^2}} \right) $.
		Then, the existing extreme theory \cite{1988AoP-Leadbetterextremal} can be directly applied.
		However, this assumption is not valid in the mixed-numerology systems due to its complicated system design.
		The extreme value theory is therefore unsuitable for this scenario.
		
		Recently, the PAPR distribution of OFDM with low subcarriers number has been studied in \cite{2018ITC-PAPR}.
		Under the condition of low subcarriers number, the mean power in the definition of PAPR (\ref{PAPR}) is no longer seen as a constant value and a ratio of two random variables are used in the PAPR analysis.
		However, the main results of \cite{2018ITC-PAPR} are derived only for the Nyquist sampled signals and the empirical approximation for continuous-time signals leads to limited precision.

		
		After revisiting previous works, we find that 
		the extreme value theory in \cite{2002-modern,2008ITWC-Derivation} and the two variables method in \cite{2018ITC-PAPR} are not suitable for the PAPR analysis of mixed-numerology transmission.
		However, the level-crossing theory that casts light on the statistical properties of signals can be exploited to extend the current method in several directions.
		First of all, the commonly used assumption that the signal is continuous and symmetric in frequency domain should be abandoned.
		Furthermore, the system parameters, such as power allocation, bandwidth, number of subcarriers can be comprehensively considered in the signal model.
		

		
	}
	
	\subsection{PAPR Analysis Based on Level-Crossing Theory}
	{\color{black}
		The level-crossing theory describes the distribution of the number of solutions to $ x(t)=r $ for a specified level $ r $.
		It has a famous formula found in \cite{1944BSTJ-RiceMathematical} for the mean number of times that the random signal crosses a specified level.
		In the following, we first discuss the probability distribution of the composite signal and then derive the PAPR distribution according to the level-crossing theory.}
	
	We rewrite the OFDM signal $ s_i(t) $ as its real and imaginary parts, 
	\begin{equation}\label{}
	s_i(t) = {x_i(t) + j\hat{x}_i(t)},
	\end{equation}
	where 
	$ \hat{x}_i(t) $ is the Hilbert transform of $ x_i(t) $ and
	$ A_k=A^R_k+jA^I_k $.
	We also suppose that $ A^R_k$ and $ A^I_k $ are uncorrelated, i.e. $ E[A^R_k A^I_k]=0 $.
	Similarly, we denote $ z(t)= x(t) + j \hat{x}(t)$, where $x(t) = \sum_{i=1}^{M} x_i(t)$ and $ \hat{x}(t) = \sum_{i=1}^{M} \hat{x}_i(t) $.
	
	According to the convergence results in \cite{2010ITIT-Convergence}, the sample sequences of $ x_i(t) $ and $ \hat{x}_i(t) $ both converge weakly to a Gaussian random process with zero mean and variance $ \eta_i/2 $ for large $ N_i $.
	Since the summation of Gaussian variables still obeys Gaussian distribution and the different subband signals are uncorrelated to each other,
	we can directly reach a conclusion that
	as $ N_i \rightarrow \infty (i=1,...,M)$, for any closed and finite interval $ T_0 $,
	\begin{equation}\label{}
	\left\{z(t), t \in T_0\right\} {\longrightarrow}{\mathcal{CN}}{(0,1)}.
	\end{equation}
	
	The normalized envelope of composite signal can be written as
	\begin{equation}\label{envelope}
	r(t) = {|z(t)|} = \sqrt{{x^2(t) + \hat{x}^2(t)}}.
	\end{equation}
	Due to the orthogonality of Hilbert transform, $ x(t) $ and $ \hat{x}(t) $ are uncorrelated Gaussian random variables for any given $ t $.
	Additionally, the uncorrelated Gaussian random variables are statistically independent.
	$ x(t) $ and $ \hat{x}(t) $ are therefore statistically independent Gaussian processes.
	Thus, we can refer to the envelope $ r(t) $ as a Rayleigh random process
	of which the probability density function (PDF) of (\ref{envelope}) is 
	\begin{equation}\label{enve_Ray}
	f_{r}(r)\rightarrow 2 r e^{-r^{2}},
	\end{equation}
	for $ N_i \rightarrow \infty \,(i=1,...,M)$.


	{\color{black}
		The PAPR distribution can be analyzed through the analysis of the peaks in random signals.}
	The probability that an arbitrary peak $ \rho $ in one composite signal exceeds the level $ r $ can be approximated by the ratio of the mean number of the peaks above the level to the mean number of total peaks,
	\begin{equation}\label{p1}
	\operatorname{Pr}\{\rho>r\}=\frac{\text{the mean number of the peaks above}\,\, r}{\text{the mean number of total peaks}}.
	\end{equation}
	Meanwhile, if we properly select a reference level $ \bar r $ so that each positive crossing of the level $ \bar r $ has a single peak exceeding the level $ r $, 
	the conditional probability of (\ref{p1}) can be approximated as \cite{2001ITC-Ochiaidistribution}
	\begin{equation}\label{pr1}
	\operatorname{Pr}(\rho>r | \rho>\overline{r})\approx
	\frac{\bar{U}(r, T_0)}{\bar{U}(\bar{r}, T_0)}, \text{for} \,\, r> \bar{r},
	\end{equation}
	where $ \bar{U}(r, T_0) $ is the mean number of points at which $ r(t) $ crosses the level $ {r} $ within the symbol duration $ T_0 $.
	
	In the next, we need to derive $ {\bar{U}({r}, T_0)} $ for the composite signals based on the level-crossing theory\cite{1944BSTJ-RiceMathematical}.
	We consider that the mixed-numerology system is designed to approximate a series of non-overlapped PSD $ G_i(f) $ ($ i=1,...,M $).
	$ G_i(f) $ is an integrable function in the interval $ [\delta_i,\delta_i+B_i] $, satisfying
	\begin{equation}\label{}
	\int_{\delta_i}^{\delta_i+B_i} G_i(f) d f=\eta_i.
	\end{equation}
	{\color{black}
		In addition, the power distribution on different subcarriers is assumed to be equal in one subband.
		\footnote{{\color{black}The case that users have different power levels within same subband can be regarded as that subband is re-divided according to users. Then, bandwidth of user can be treated as an individual subband.}}
		We only consider the power allocation on different subbands.}
	Then we have $ G_i(f)= \eta_i/B_i$ within the interval $ [\delta_i,\delta_i+B_i] $.
	Obviously, $ G_i(f)\!\cdot\! f $ and  $ G_i(f)\!\cdot\! f^2 $ are also integrable.
	Thus, the first and the second normalized spectral moments of $ s_i(t) $ can be calculated as
	\begin{equation}\label{moment1}
	\alpha_i = 
	\dfrac{1}{\eta_i}\int_{\delta_i}^{\delta_i+B_i} 2\pi f G_i(f) d f
	= {2\pi}\left( \delta_i+\frac{B_i}{2}\right),
	\end{equation}
	\begin{equation}\label{moment2}
	\begin{split}
	\beta_i &= 
	\\
	&\dfrac{1}{\eta_i}\int_{\delta_i}^{\delta_i+B_i}(2\pi f)^2 G_i(f) d f
	= {4\pi^2} \left( \delta_i^2+ \delta_i B_i + \frac{1}{3} B_i^2\right).
	\end{split}
	\end{equation}
	{\color{black}
		From (\ref{moment1}) and (\ref{moment2}) we find that if the OFDM signals is symmetric and continuous in frequency domain,
		the first spectral moment of the signal turns to be zero and the second spectral moment has a simple form $ \dfrac{\pi^2}{3}B_i^2 $, which leads to a special case in \cite{2001ITC-Ochiaidistribution}.
		Different to previous works,
		it can be seen from Fig. \ref{psd} that the signal model of mixed-numerology has a more general form.
		In order to consider its complicated system design, the spectral moments of all subbands should be particularly expressed.
	}

	According to level-crossing theory, the mean number of level-crossing points can be calculated as \cite{1944BSTJ-RiceMathematical}
	\begin{equation}\label{Rice}
	\bar{U}(r, T_0) = {1\over 2 }T_0 f_{r}(r) E[ | \dot{r}(t) \| r(t)=r],
	\end{equation}
	{\color{black}
		where $ f_{r}(r) $ is the probability density function of the envelope $ r(t) $ and $ E[ | \dot{r}(t) \| r(t)=r] $ denotes the conditional expectation of derivative of $ r(t) $.}
	Then, the analytical expression of (\ref{Rice}) can be derived in Appendix as
	\begin{equation}\label{equ1}
	\bar{U}(r, T_0) =  \sqrt{{\Lambda}}\cdot r  \exp \left({-r^2}\right),
	\end{equation}
	where 
	\begin{equation}\label{equ2}
	\begin{split}
	\Lambda &{ \buildrel \Delta \over =}    {{T_0^2\over \pi}\left(\sum_{i=1}^{M}{\beta _i}\eta_i - \left({ \sum_{i=1}^{M}\alpha _i\eta_i}\right)^2\right)} \\
	&=  
	{
		{\pi} \mu^2  \sum_{i=1}^{M} n_i^2[{1\over 3}\eta_i^2N_i^2 + 4\eta_i(1\!-\!\eta_i)(d_i^2 + d_iN_i +{1\over 3}N_i^2)]
	}\\
	& {\color{black}\quad - 8\pi\mu^2\!\!\sum_{m=1}^{M}\!\sum_{l=1,l\neq m}^{M}\!\!\!\!\left[   n_ln_m\eta_l\eta_m(d_l + {1\over 2}N_l)(d_m + {1\over 2}N_m)\right]\!}
	.
	\end{split}
	\end{equation}
	Here we denote $ d_i = \delta_i / f_i $ as the normalized frequency shift
	{\color{black}and $ \mu = T_0/T_{sys,1} $}.
	
	Substituting (\ref{equ1}) into (\ref{pr1}), we have
	\begin{equation}\label{}
	\operatorname{Pr}(\rho>r | \rho>\bar{r})=
	\frac{r e^{-r^{2}}}{\bar{r} e^{-\bar{r}^{2}}},
	\end{equation}
	and accordingly
	\begin{equation}\label{}
	\operatorname{Pr}(\rho<r | \rho>\bar{r})= 1-
	\frac{r e^{-r^{2}}}{\bar{r} e^{-\bar{r}^{2}}}.
	\end{equation}
	
	{\color{black}
		From the probability distribution of one peak, the PAPR distribution can be derived by considering all the peaks in one composite signal.
		The mean number of peaks beyond the reference level $ \bar r $ within symbol duration is $ {\bar{U}(\bar{r}, T_0)} $. 
		Meanwhile, we assume that all the peaks are uncorrelated statistically.
		In fact, this assumption is weakened when multiple CP parts are observed over the LCM duration.
		However, from Fig. \ref{symbol} we can see that as the composite signal is composed of multiple signals from independent subbands, 
		peaks are also accumulated by different subbands and the extended symbol duration is not a full repetition of any part of the LCM symbol.
		It means that the correlation between the beginning and the following peaks is reduced. 
		As more numerologies adopted, the correlation can be further diminished. 
		Furthermore, since CP only takes a small percentage of the symbol duration, most peaks in LCM duration are still uncorrelated to each other.
		We therefore keep this assumption in the following derivation.
	}
	
	Then, the conditional cumulative distribution function (CDF) of the envelope can be calculated as
	\begin{equation}\label{}
	F(r | r>\bar{r})=\operatorname{Pr}(\rho<r | \rho>\bar{r})^{\bar{U}(\bar{r}, T_0)}.
	\end{equation}
	Additionally, $ \bar r $ is normally set to make all peaks larger than the level $ \bar r $,
	which implies $ F\left( {\bar r} \right)= \operatorname{Pr}(r<\bar{r})\approx 0$.
	Consequently, the CDF of the envelope of composite signal can be simply expressed as
	\begin{equation}\label{cdf}
	\begin{split}
	F(r) &{\color{black}= F\left( {\bar r} \right) + F(r|r > \bar r)\left( {1 - F\left( {\bar r} \right)} \right) \approx F(r | r>\bar{r})} \\
	&{\color{black}=  \left(1-\frac{r e^{-r^{2}}}{\overline{r} e^{-\overline{r}^{2}}}\right)^{\bar{U}(\bar{r}, T_0)},}
	\end{split}
	\end{equation}
	for $ r>\bar{r} $.
	
	For large $ r $, according to the limiting form of the exponential function, we can further simplify the CDF in Eq. (\ref{cdf}) as
	\begin{equation}\label{proposedCDF}
	\begin{split}
	{\color{black}{\left( {1 - \frac{{r{e^{ - {r^2}}}}}{{\bar r{e^{ - {{\bar r}^2}}}}}} \right)^{\bar U(\bar r,{T_0})}} }
	&{\color{black}= {\left( {{{\left( {1 - \frac{{r{e^{ - {r^2}}}}}{{\bar r{e^{ - {{\bar r}^2}}}}}} \right)}^{ - \frac{{\bar r{e^{ - {{\bar r}^2}}}}}{{r{e^{ - {r^2}}}}}}}} \right)^{ - \sqrt \Lambda   \cdot r{e^{ - {r^2}}}}}} \\
	&{\color{black}= \exp \left( { - \sqrt \Lambda   \cdot r{e^{ - {r^2}}}} \right)}.
	\end{split}
	\end{equation}
	The corresponding CCDF of PAPR is therefore obtained as
	\begin{equation}\label{ccdf}
	{\rm{CCDF}}(\gamma) \approx 1 - \exp \left(- \sqrt{{\Lambda}\!\cdot \!\gamma}\, e^{-\gamma}\right),
	\end{equation}
	which is not dependent upon the reference level $ \bar{r} $.
	{\color{black}
		We can find that the proposed Eq. (\ref{ccdf}) takes the system parameters into full consideration, including numerology, bandwidth and power level of each subband.
		In addition, if only one numerology is considered, it leads to 
		the same result as (\ref{R1}).
	}
	
	
	
	{\it{Remark 2:}}
	{\color{black}
		In fact, the derived result is not limited to the generalized synchronized system.
		When there is no integral LCM of the signal periods for each subband, 
		which makes (\ref{period}) invalid and is referred to as the asynchronous  system\cite{2017ICM-ZhangMulti},
		the asynchronous signals is still a complex Gaussian process. 
		Thus, the proposed expression (\ref{ccdf}) can be easily applied to this case as will be shown in simulation. 
	}
	On the other hand, when we let $ f_i=f_j $ for any $ i,j=1,...,M $, Eq. (\ref{ccdf}) turns to be the CCDF of PAPR for NC-OFDM system.
	As the existing PAPR analysis of NC-OFDM lacks of theoretical results,
	the proposed analytical expression (\ref{ccdf}) can fill up this gap.
	
	{\color{black}
		{\it{Remark 3:}}
		Filtering and windowing are two main spectrum confinement techniques to reduce the out-of-band emissions as well as the interference from different numerologies in 5G NR\cite{2016ICM-Waveform,2016ICM-ZhangWaveform}.
		Filtered-OFDM (F-OFDM) is a candidate waveform that applys a digital filter with predesigned frequency response\cite{2016ICM-ZhangWaveform}.
		Windowing is to multiply the samples at the edges of the symbol by raised-cosine coefficients, which is known as W-OFDM\cite{2016ICM-Waveform}.
		Both techniques change the envelope of composite signals that will not be a strict Rayleigh process.
		Hence, the analysis of PAPR distribution can be very difficult due to the intricacy of the probability density of the filtered/windowed signal.
		Actually, applying windowing technique only changes the samples at the symbol edges, which generally has little or no PAPR overhead\cite{2017ICM-Lien5G}.
		Meanwhile, filtering process makes excess time-spread of each symbol due to the filter tailing.
		As the tailing is still contained in the successive symbol during the frame processing, we calculate the PAPR of filtered signal over the original symbol duration $ T_0 $ for fair comparison.
		With limited changes on signal envelope,
		filtering also has small influence on PAPR, as will be shown in the simulation.
		Therefore, the derived expressions can still serve as CCDF approximation for F-OFDM and W-OFDM.
	}

	\subsection{Relationship Between PAPR and Power Allocation}
	In mixed-numerology system, power allocation are normally addressed to maximize the information sum rate of multi-user, according to the instantaneous channel state information (CSI) available at the transmitter or the QoS requirements of the users.
	From Eq. (\ref{equ2}), we know that the power allocation among different subbands also has influence on PAPR distribution.
	
	To evaluate the PAPR under different power allocation schemes, we take into account of the mean envelope, which can be calculated as
	\begin{equation}\label{PAPR_mean}
	\begin{split}
	&E[r] = \int_0^\infty  r  dF\left( r  \right) \\
	&= {\int_0^\infty  {\sqrt \Lambda  r\left( {2{r^2} - 1} \right){e^{ - {r^2}}}\left( {1 - \frac{{r{e^{ - {r^2}}}}}{{\bar r{e^{ - {{\bar r}^2}}}}}} \right)} ^{\sqrt \Lambda  \bar r{e^{ - {{\bar r}^2}}} - 1}}dr.
	\end{split}
	\end{equation}
	Apparently, it is difficult to deal with this expression,
	since the integral of (\ref{PAPR_mean}) has to be carried out numerically.
	
	Instead, we may focus on $ \sqrt \Lambda $ that contains entire system parameters.
	In the following, the derivative of (\ref{PAPR_mean}) with respect to $ \sqrt{\Lambda} $ is considered.
	Because of the uniform convergence of the PDF, the derivative of the integral is dealt with by the derivative of the integrand of (\ref{PAPR_mean}), which is given in (\ref{derive_mean}) at the top of next page.
	\newcounter{tem}
	\begin{figure*}
		\normalsize
		\setcounter{tem}
		{\value{equation}}
		
		\begin{equation}\label{derive_mean}
		\begin{array}{l}
		\frac{d}{{d\sqrt \Lambda  }}E\left[ r \right] = \int_0^\infty  {\frac{d}{{d\sqrt \Lambda  }}\left[ {\sqrt \Lambda  r\left( {2{r^2} - 1} \right){e^{ - {r^2}}}{{\left( {1 - \frac{{r{e^{ - {r^2}}}}}{{\bar r{e^{ - {{\bar r}^2}}}}}} \right)}^{\sqrt \Lambda  \bar r{e^{ - {{\bar r}^2}}} - 1}}} \right]} dr\\
		= \int_0^\infty  {r\left( {2{r^2} - 1} \right){e^{ - {r^2}}}\left[ {{{\left( {1 - \frac{{r{e^{ - {r^2}}}}}{{\bar r{e^{ - {{\bar r}^2}}}}}} \right)}^{\sqrt \Lambda  \bar r{e^{ - {{\bar r}^2}}} - 1}} + \sqrt \Lambda  \bar r{e^{ - {{\bar r}^2}}}\left( {\sqrt \Lambda  \bar r{e^{ - {{\bar r}^2}}} - 1} \right){{\left( {1 - \frac{{r{e^{ - {r^2}}}}}{{\bar r{e^{ - {{\bar r}^2}}}}}} \right)}^{\sqrt \Lambda  \bar r{e^{ - {{\bar r}^2}}} - 2}}} \right]} dr\\
		\ge 
		\left[ {\sqrt \Lambda  \bar r{e^{ - {{\bar r}^2}}}\left( {\sqrt \Lambda  \bar r{e^{ - {{\bar r}^2}}} - 1} \right) + \frac{1}{{\sqrt \Lambda  }}} \right]E\left[ r \right]
		\end{array}
		\end{equation}
		\hrule
		\vspace*{4pt}
	\end{figure*}
	
	Since $ N_i $ is normally large for $ i =1,...,M$, we have $ \sqrt \Lambda  \bar r{e^{ - {{\bar r}^2}}} \gg 1 $.
	Then, it can be seen from (\ref{derive_mean}) that
	\begin{equation}\label{}
	\frac{d}{{d\sqrt \Lambda  }}E\left[ r \right] > 0,
	\end{equation}
	{\color{black}
		which shows that the mean envelope is proportional to $ \Lambda $.}
	Therefore, the relationship between PAPR and power allocation can be reflected through a function of subbands powers, $ \Lambda(\eta_1,\eta_2,...,\eta_M) $.
	
	{\color{black}
		Next, we consider the minimization and maximization of mean envelope. 
		Intuitively, the PAPR will be reduced as the decrease of subcarrier numbers.
		Therefore, 
		when the system bandwidth is divided into several subbands according to the service requirements,
		the minimal mean envelope can be reached as long as the power is concentrated on the subband that has the smallest subcarrier number and the other subbands with more subcarriers remain idle.
		It is straightforward for the power allocation scheme to reach its minimal value. }
	
	{\color{black}
		On the other hand, the maximal mean envelope can be considered in the system design.
		When the division of system bandwidth is fixed, we can theoretically obtain the maximal mean envelope.
		Then, the output back-off of PA can be properly determined.
	}
	In order to find the theoretical maximum of mean envelope, we then take the negative to $ \Lambda $ and formulate the following problem,
	\begin{subequations}
		\label{PAPR_model}
		\begin{align}\label{sub,a}
		\mathop {\min }\limits_{\bm{\eta }=[\eta_1,...,\eta_M]^T}  &- \frac{\pi}{T_0^2}\Lambda \left( {\bm{\eta }} \right) = \dfrac{1}{2}{{\bm{\eta }}^T}{\bf{P{\bm{\eta }} }} + {{\bm{q}}^T{\bm{\eta } }},\\
		s. \, t. \quad& {{\bf{1}}^T{\bm\eta }} = 1,
		\label{sub,b}\\
		&\eta_i > 0  \quad \text{for}\quad \forall i=1,2...,M.
		\end{align}
	\end{subequations}
	where 
	\begin{equation}\label{}
	\begin{array}{l}
	{\bf{P}} = 2\left[ {\begin{array}{*{20}{c}}
		{ \alpha _1^2}&\alpha _1\alpha _2& \cdots &\alpha _1\alpha _M\\
		\alpha _1\alpha _2&{ \alpha _2^2}& \cdots &\alpha _2\alpha _M\\
		\vdots &\vdots& \ddots & \vdots \\
		\alpha _1\alpha _M&\alpha _2\alpha _M& \cdots &{ \alpha _M^2}
		\end{array}} \right]\\
	\end{array}
	\end{equation}
	and 
	\begin{equation}\label{}
	\begin{array}{l}
	{\bm{q}} = {\left[ {\begin{array}{*{20}{c}}
			{{-\beta _1}}&{{-\beta _2}}& \cdots &{{-\beta _M}}
			\end{array}} \right]^T}.
	\end{array}
	\end{equation}
	Obviously, we can find that $ \rm{rank}({\bf{P}})=1 $ and the nonzero eigenvalue of $ {\bf{P}} $ is $ 2\sum_{i=1}^{M}\alpha_i^2 $.
	Since $ {\bf{P}} $ is semi-definite, this problem (\ref{PAPR_model}) is a second-order programing and can be solved by classical optimization algorithms\cite{2004-BoydConvex}.
	
	Particularly, when $ M=2 $, the problem (\ref{PAPR_model}) is
	equivalent to solving the following set of 3 linear equations \cite{2004-BoydConvex}, i.e.,
	\begin{equation}\label{equ_set}
	\left[ {\begin{array}{*{20}{c}}
		{\bf{P}}&{\bf{1}}_M\\
		{{{\bf{1}}_M^T}}&0
		\end{array}} \right]\left[ {\begin{array}{*{20}{c}}
		{{{\bm{\eta }}^ * }}\\
		{{\nu ^*}}
		\end{array}} \right] = \left[ {\begin{array}{*{20}{c}}
		{ - {\bm{q}}}\\
		{{1}}
		\end{array}} \right],
	\end{equation}
	where $ \nu ^* $ is the Lagrangian multiplier.
	Denote $ {\bf{A}} = \left[ {\begin{array}{*{20}{c}}
		{\bf{P}}&{\bf{1}}_M\\
		{{{\bf{1}}_M^T}}&0
		\end{array}} \right] $ as the Karush-Kuhn-Tucker (KKT) matrix of (\ref{PAPR_model}).
	We can find that $ \rm{rank}({\bf{A}})=3 $.
	Thus, the KKT matrix is non-singular so that the problem (\ref{PAPR_model}) has the unique solution 
	\begin{equation}\label{solution}
	\begin{split}
	&\eta_1^* = \frac{1}{2(\alpha_1-\alpha_2)^2}(\beta_1-\beta_2+2\alpha_2^2-2\alpha_1\alpha_2), \\
	&\eta_2^* = \frac{1}{2(\alpha_1-\alpha_2)^2}(\beta_2-\beta_1+2\alpha_1^2-2\alpha_1\alpha_2).
	\end{split}
	\end{equation}
	More specially, when two subbands have the same bandwidth and the guard band is approximated to zero, ($ B_1 = B_2$ and $ G_1 \approx0 $),
	it can be calculated that $ \eta_1^* = \eta_2^* = 1/2 $.
	
	{\it{Remark 4:}}
	{\color{black}
		From this result, we find that it is the bandwidth of subband rather than subcarrier numbers takes more important role in PAPR distribution.
	}
	As the subcarrier spacing in the first subband is smaller than the second subband,
	the first subband possesses more subcarriers when they have equal bandwidths.
	The optimal solution indicates that the power is evenly distributed in the system bandwidth regardless of the numerology selection of the second subband.

	

	\section{Numerical Results}\label{sec5}
	In this section, simulations are performed to validate the proposed analytical expressions in mixed-numerology systems.
	In the subsequent simulations, $ 10^6 $ LCM symbols are randomly generated with 16-QAM modulation, the guard bands are set as $ G_i=20f_1 $ for any two adjacent subbands
	{\color{black} and the CP length is set as $ 7\% $ of the symbol period $ T_{sys,i} $.}
	The total power is equally allocated to each subband if not specified otherwise.
	
	{\color{black}
		To evaluate the PAPR distribution for practical scenario, 
		the continuous-time signals are considered,
		which can be accurately approximated by $ J $-time oversampling in the discrete time domain when $ J\geq4 $ \cite{2001ICL-TellamburaComputation}.
		In the following simulations, we use $ 8 $-time oversampled signals to evaluate the PAPR of composite signals.
	}
	
	\subsection{CCDF of PAPR for Mixed-Numerology Transmission}
	

	\begin{figure}[t]
		\centering
		\includegraphics[width=88mm]{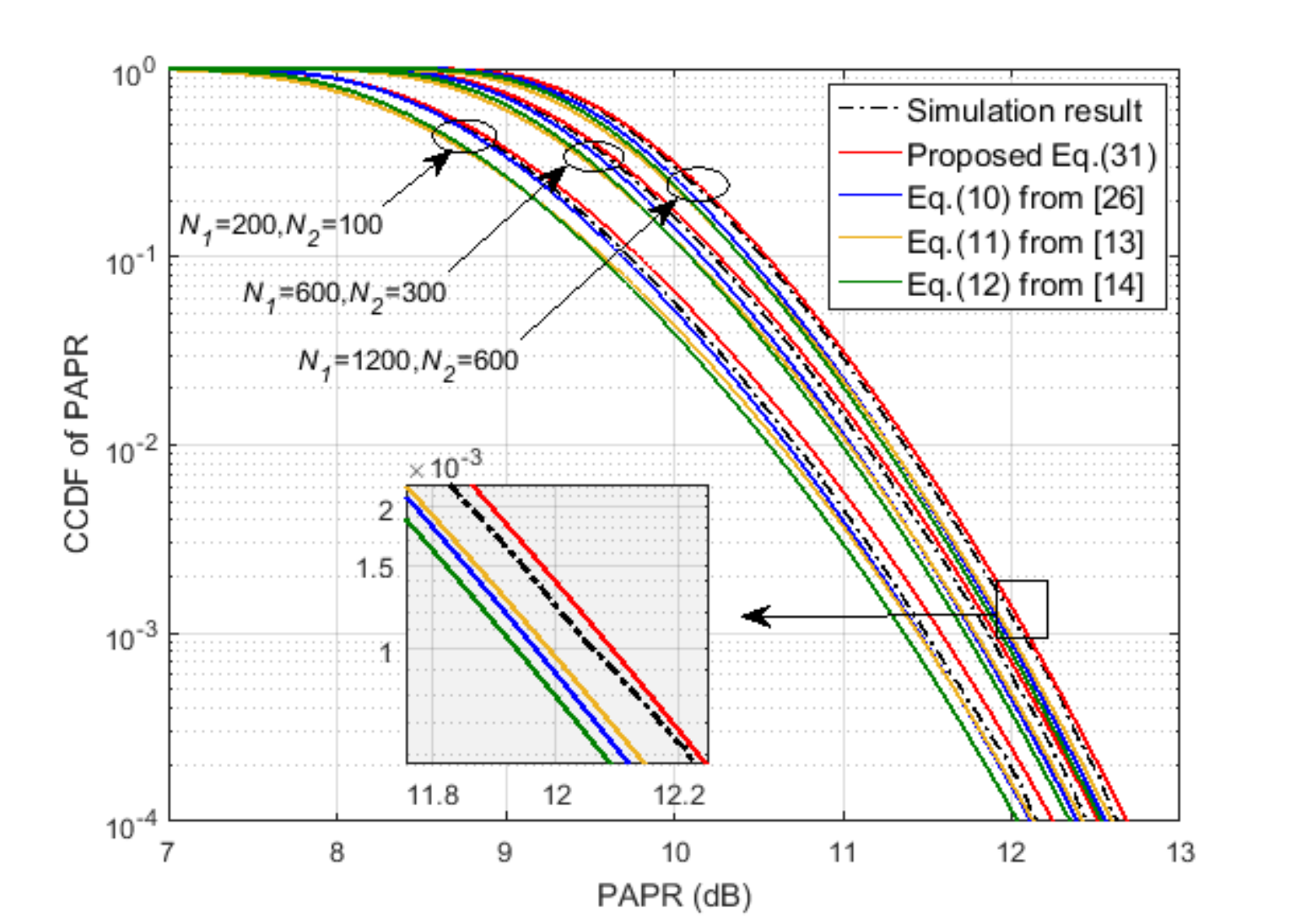}
		\caption{{\color{black}Comparison of analytical and simulation results of the CCDF of PAPR for two numerologies.}}
		\label{numero2}
	\end{figure}
	Fig. \ref{numero2} shows the results in the two numerologies system with the subcarrier spacings satisfying $ f_2=2f_1 $.
	The numbers of subcarriers for each subband are set as $ ``N_1=200,N_2=100", ``N_1=600,N_2=300", ``N_1=1200,N_2=600" $, respectively.
	{\color{black}
		We also compare the proposed expression with previous works, including (\ref{emp}) in \cite{1998Nee}, (\ref{R1}) in \cite{2001ITC-Ochiaidistribution} and (\ref{R2}) in \cite{2002-modern}, in which the number of subcarriers are calculated as the sum of subcarrier numbers of all subbands, i.e., $ N=N_1+N_2 $.
		It can be seen that the CCDF curves of empirical expression (\ref{emp}) show discernible deviation from the simulation results when employing large numbers of subcarriers.
		In addition, the existing theoretical results also fail to match with the simulation results.
	}
	However, the proposed analytical expression (\ref{ccdf}) achieves better approximation as the numbers of subcarriers increase.
	The accuracy of the results also implies that the composite signal weakly converges to the Gaussian process as $ N_i $'s increase.
	
	\begin{figure}[t]
		\centering
		\includegraphics[width=88mm]{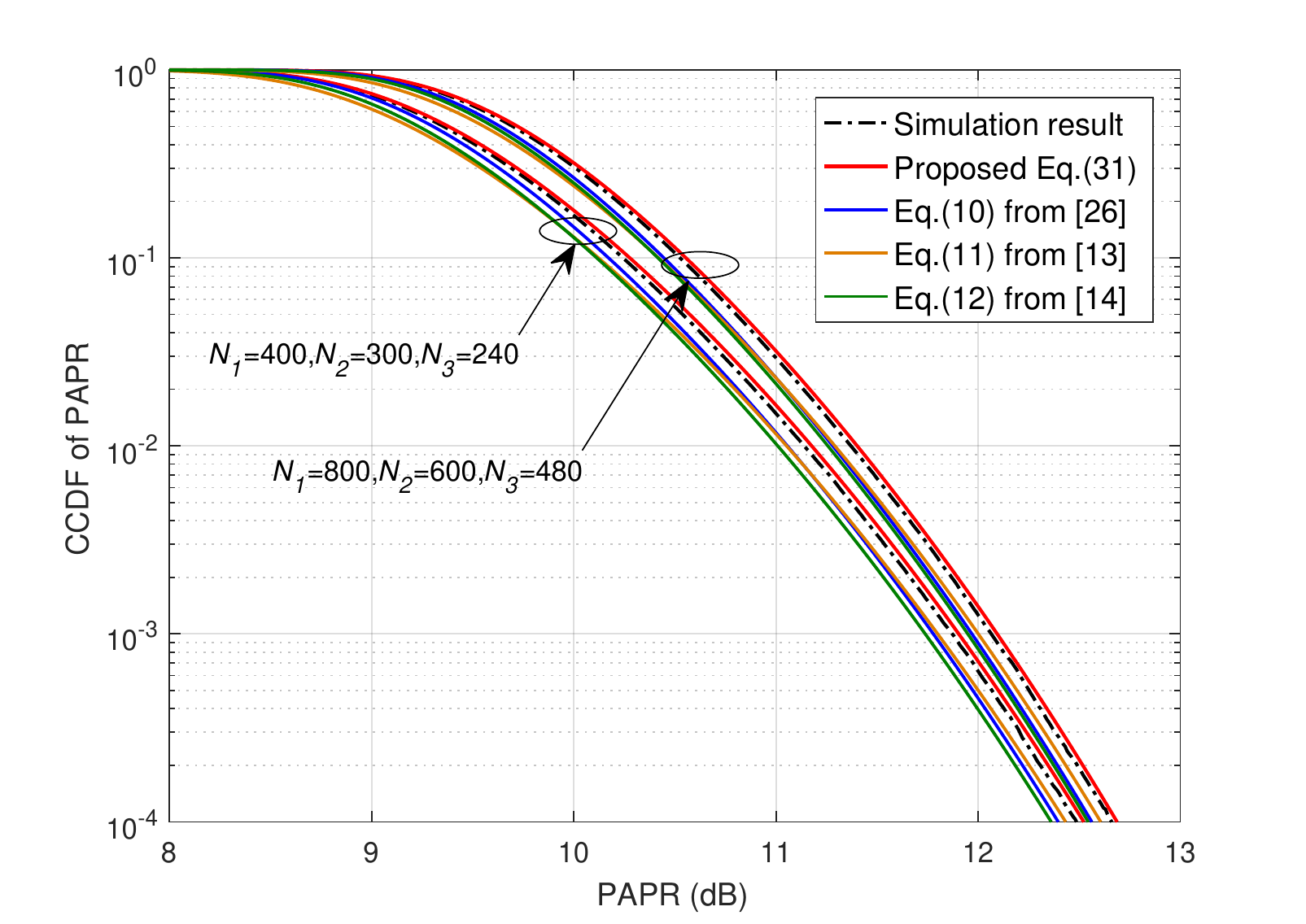}
		\caption{{\color{black}Comparison of analytical and simulation results of the CCDF of PAPR for asynchronous system with three numerologies.}}
		\label{numero3}
	\end{figure}
	We further perform the simulation for asynchronous system using three numerologies with subcarrier spacings $ 5f_1=4f_2=3f_3 $.
	In this case, the symbol period of subbands OFDM signal are fractional multiple of the duration of composite signal, i.e., $ n_2=0.8n_1, n_3= 0.6n_1$, {\color{black}and the starting point is random for every frame.
		The numbers of subcarriers are chosen as $ ``N_1=400,N_2=300,N_3=240"$, and $ ``N_1=800,N_2=600,N_3=480"$, respectively.}
	It can be observed that the existing works lead to obvious deviation from the simulation results.
	As expected, 
	both the simulation and proposed approximation results are nearly overlapped when the subcarrier numbers become large.

	\begin{figure}[t]
		\centering
		\includegraphics[width=88mm]{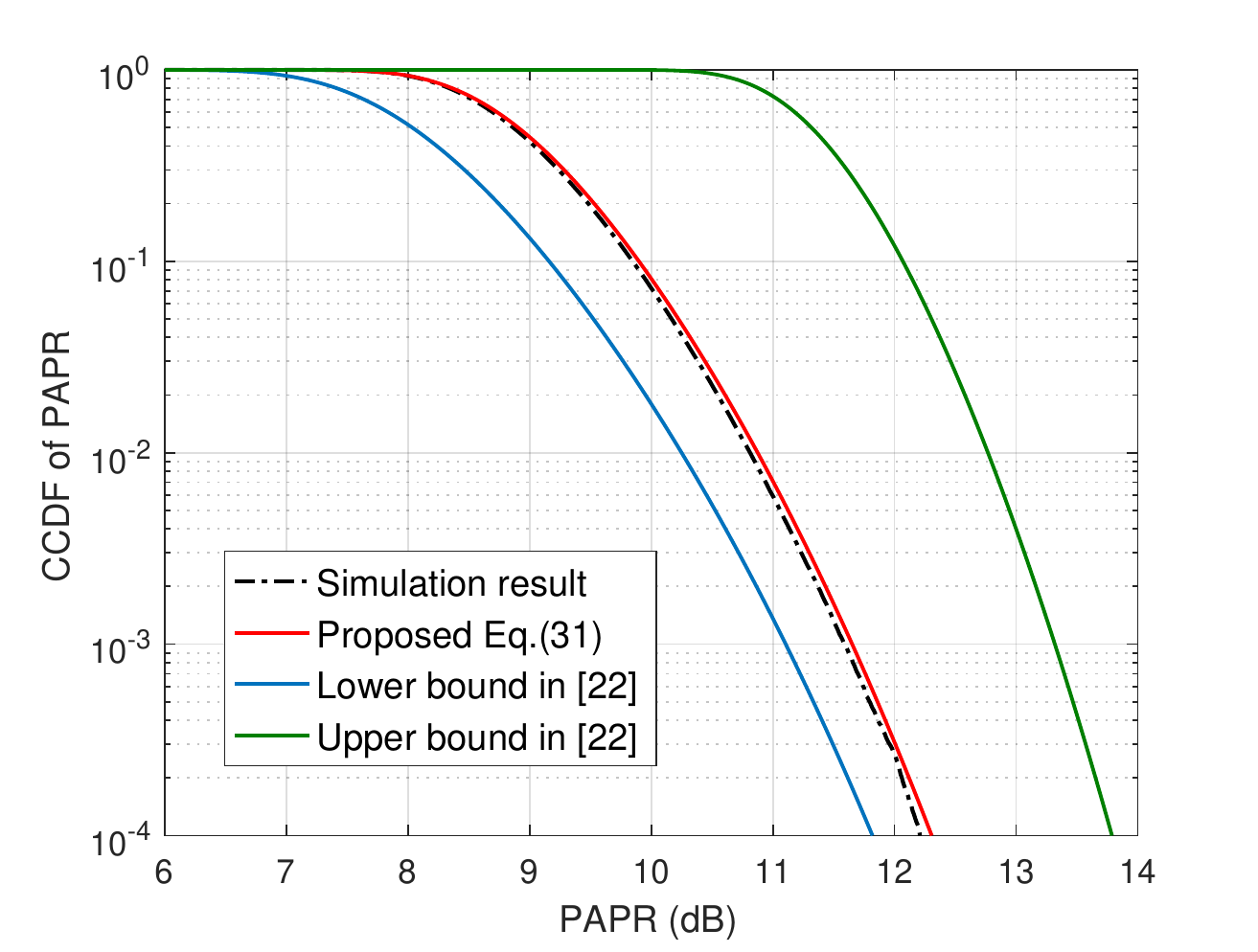}
		\caption{{\color{black}Comparison of approximations and simulation results of the CCDF of PAPR for NC-OFDM with two subbands.}}
		\label{NC1}
	\end{figure}
	{\color{black}
		As for NC-OFDM system, we compare the simulated results of NC-OFDM system with the proposed analytical expression (\ref{ccdf}), as shown in Fig. \ref{NC1}.
		The lower and upper bounds derived in \cite{2017ToETT-PAPR} are also plotted for comparison.
		We set the total number of subcarriers as 512 and two subbands occupy 200 subcarriers, respectively.
		The spectrum notch equals to 112 subcarrier spacings.
		We can observe that the previous CCDF bounds are far away from the simulated curve and the proposed expression (\ref{ccdf}) is very closed to the simulation result.
	}
	
	\begin{figure}[t]
		\centering
		\includegraphics[width=88mm]{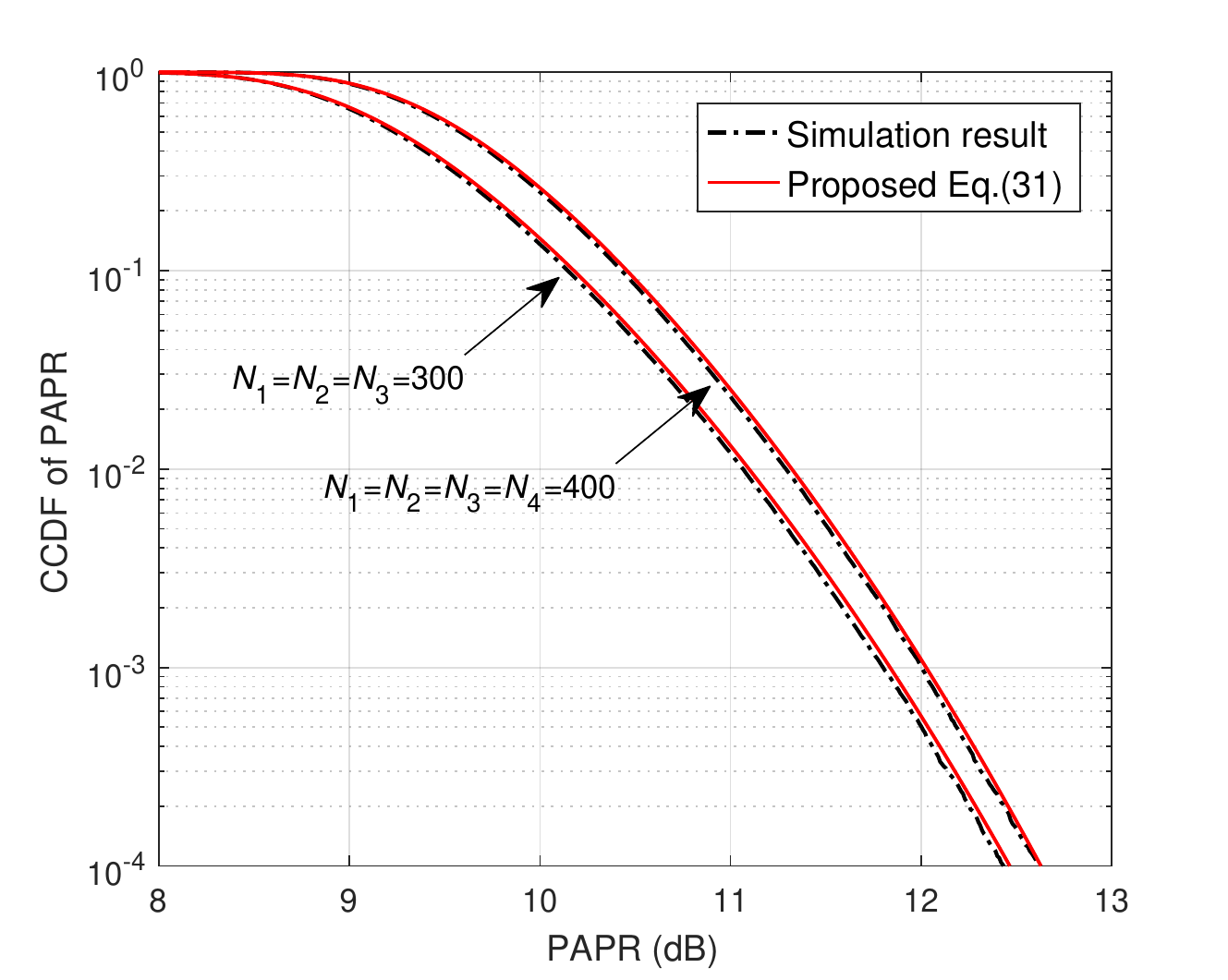}
		\caption{{\color{black}Comparison of approximations and simulation results of the CCDF of PAPR for NC-OFDM with three and four subbands.}}
		\label{NC2}
	\end{figure}
	{\color{black}
		We then set more subcarriers and more spectrum divisions for NC-OFDM.}
	The total number of subcarriers are set as 1024 and 2048, respectively.
	The system bandwidths are divided into three and four subbands each of which has 300 and 500 continuous data subcarriers to carry information symbols and 62 and 16 null subcarriers to serve as 
	{\color{black}
		spectrum notches.
		Since the derived bounds in \cite{2017ToETT-PAPR} is merely for the case of two subbands, we only compare the simulation results with our proposed expression.}
	From Fig. \ref{NC2}, it can be seen that the curves of proposed expression closely follow the simulation results.
	This demonstrates that the proposed expression has good generality for the OFDM-based signals with non-continuous frequency spectrum.

	\begin{figure}[t]
		\centering
		\includegraphics[width=88mm]{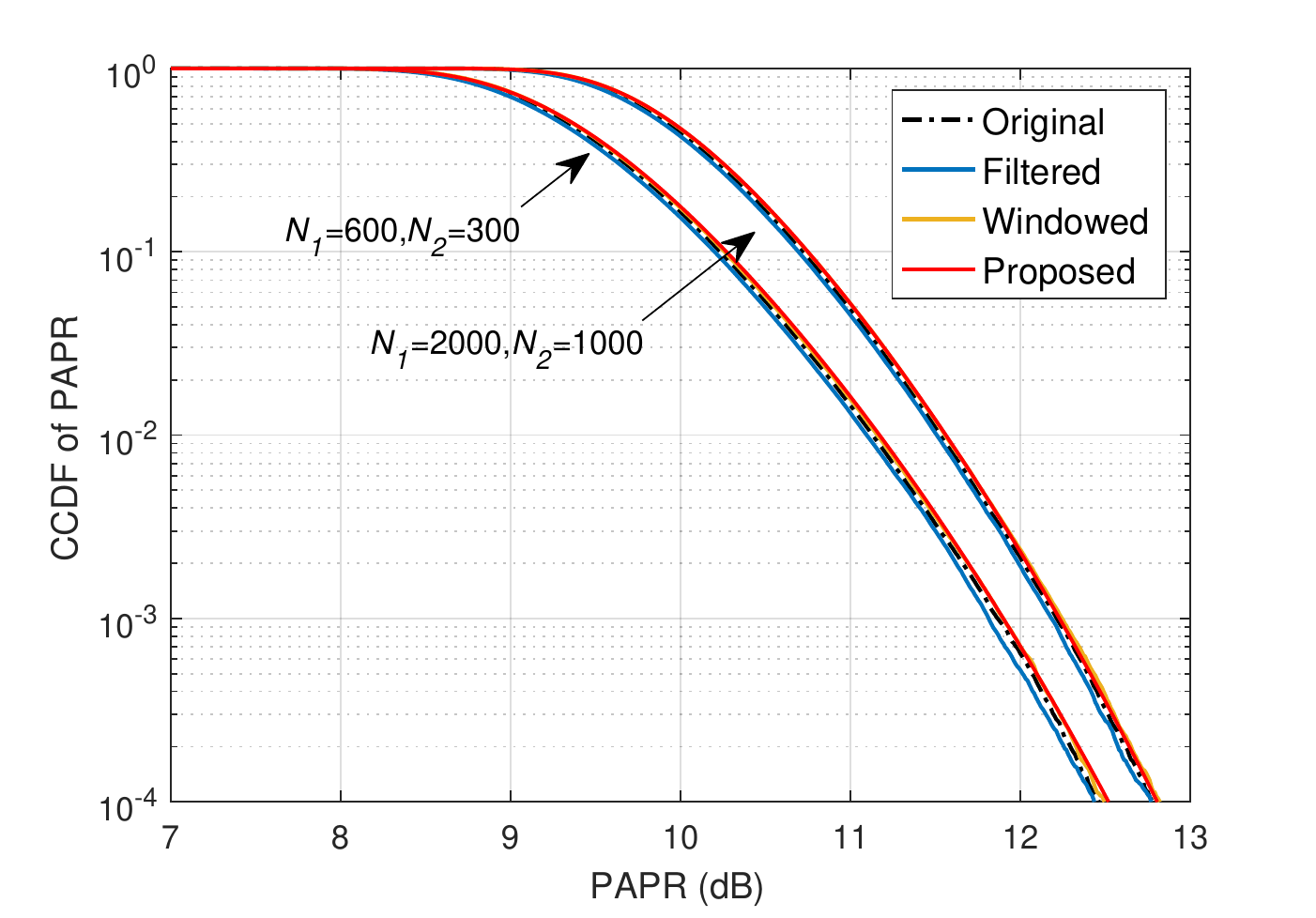}
		\caption{{\color{black}Comparison of analytical and simulation results of the CCDF of PAPR with filtering and windowing for two numerologies.}}
		\label{FandW}
	\end{figure}
	{\color{black}
		We also consider the PAPR distribution of F-OFDM and W-OFDM in Fig. \ref{FandW}.
		Subband-specific filter is designed based on the truncated Sinc finite impulse response (FIR) with 512 order for F-OFDM\cite{2016ICM-ZhangWaveform,2017IJoSAiC-Guan5G}.
		W-OFDM adopts raised cosine window suggested by \cite{2016ICM-Waveform}.  
		As expected, both F-OFDM and W-OFDM achieve similar PAPR distribution to the original OFDM.
		The proposed approximation still provide good match to the simulation results.
	}
	
	\subsection{Effect of Power Allocation on PAPR}

	\begin{figure}[t]
		\centering
		\includegraphics[width=88mm]{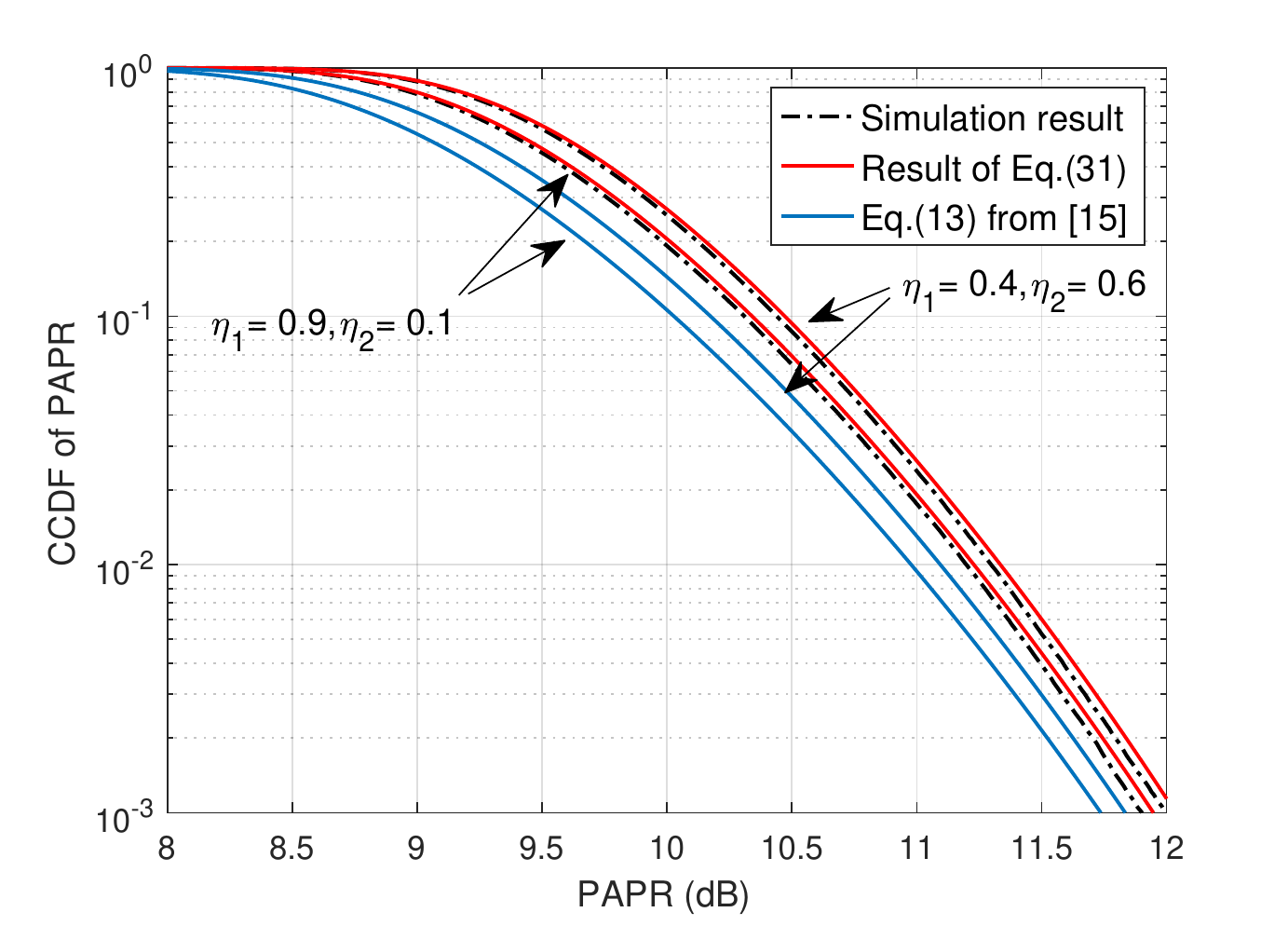}
		\caption{{\color{black}Comparison of analytical and simulation results of the CCDF of PAPR with different power allocation for two numerologies.}}
		\label{ccdf_power}
	\end{figure}
	We now study the PAPR distribution under the different power allocation schemes.
	Fig. \ref{ccdf_power} shows the CCDF of PAPR for two numerologies systems with $ f_2=2f_1 $ and $ N_1=1000,N_2=500 $.
	The power on the first and second subbands are set as $ \eta_1=0.9,\eta_2=0.1 $ and {\color{black}$ \eta_1=0.4,\eta_2=0.6 $}, respectively.
	We also plot the result of (\ref{R3}) for comparison.
	It can be seen that the proposed expression still fits the simulation results.
	{\color{black}
		Furthermore,
		when two subbands have similar power allocation,
		the curve of CCDF of $ \eta_1=0.4,\eta_2=0.6 $ is on the rightmost.}
	Compared with the case that most power is concentrated on the first subband, it probably leads to higher PAPR when the power is equally distributed to the two subbands, as analyzed in the previous section.

	\begin{figure}[t]
		\centering
		\includegraphics[width=88mm]{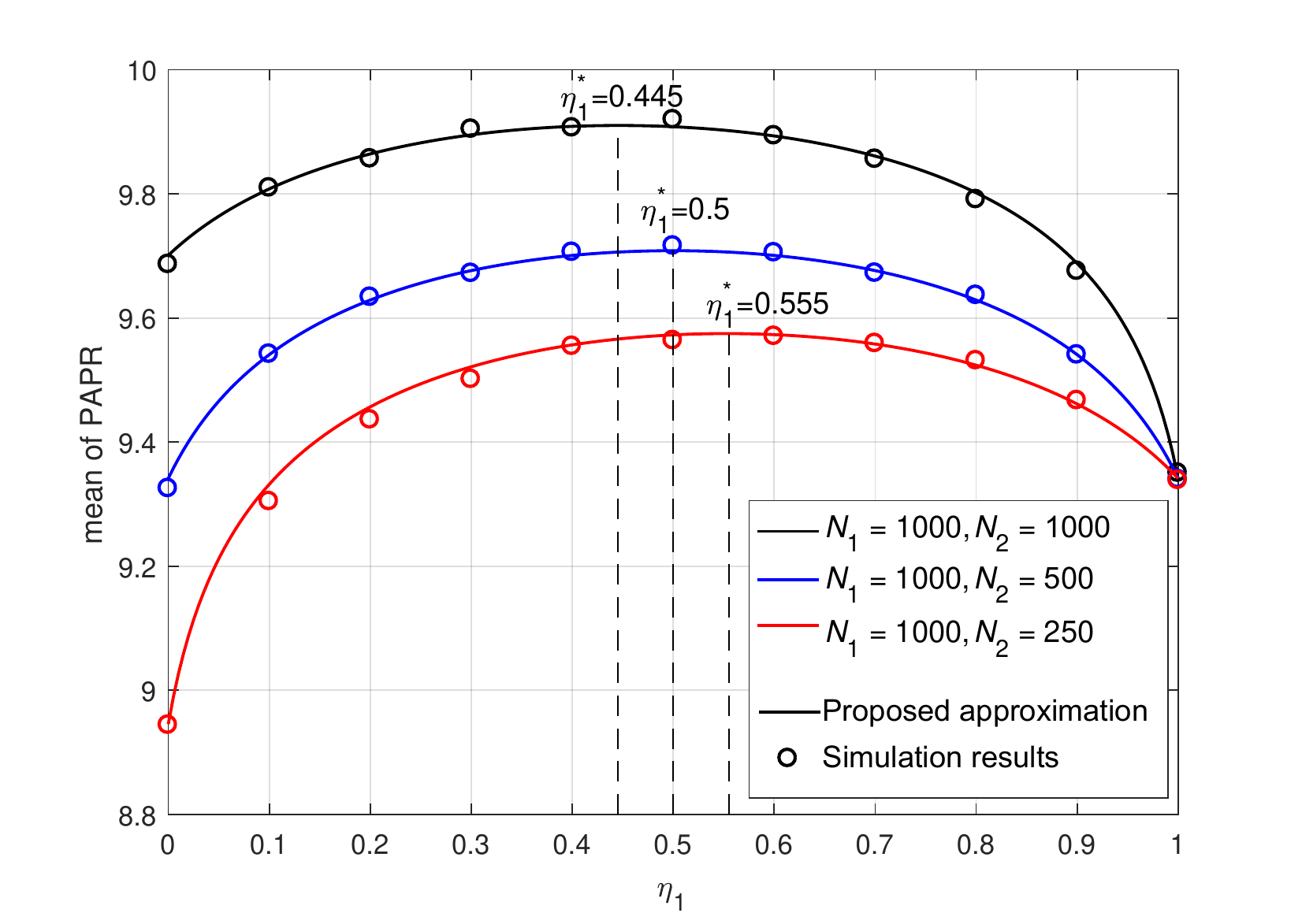}
		\caption{Mean of PAPR for two numerologies systems with different system bandwidths.}
		\label{meanPAPR}
	\end{figure}
	To analyze the PAPR influence of power allocation to different subbands,
	we numerically calculate the mean of PAPR using (\ref{PAPR_mean}).
	In Fig. \ref{meanPAPR}, a two numerologies system with $ 2f_1=f_2 $ is considered and 
	the ratio of the first subband bandwidth to the second subband bandwidth $ {B_1}/{B_2} $ is set as $ 1:2, 1:1, 2:1 $, respectively.
	Note that the horizontal coordinate is $ \eta_1 $ from 0 to 1, denoting the percentage of total power in the first subband and the power allocation to the second subband is correspondingly $ \eta_2=1-\eta_1 $.
	The calculated solution using (\ref{solution}) is also marked by the dashed line.
	The simulation results is also dotted as the reference.
	It is observed that the closed-form solution $ \eta^* $ can always reach the maximum PAPR which proves the optimality of (\ref{solution}) for the problem (\ref{PAPR_model}).
	Moreover, 
	when $ N_1=1000,N_2=500 $, which means that each subband has the same bandwidth, the maximal mean PAPR is occurred at equal power allocation despite that the first subband has more subcarriers.
	Additionally,
	as the second subband improves the bandwidth,
	the maximal PAPR will be achieved when more power is allocated to the second subband.
	It can be explained that the subband with with larger bandwidth has stronger effect on PAPR.
	In addition, the curves of mean PAPR show an almost flat transition as $ \eta_1 $ increases from 0.3 to 0.7.
	It indicates that PAPR is not sensitive to the power fluctuation when the subbands bandwidths are fixed.

	\section{Conclusions}
	{\color{black}
		In this paper, we proposed an analytical expression of PAPR distribution for mixed-numerology system.
		The level-crossing theory is reconsidered to develop a new expression to approximate the CCDF of PAPR, 
		which can be applied to not only the mixed-numerology system but also the NC-OFDM system.
		In addition, we investigated the effect of power allocation on PAPR.
		The derived result indicates that the subbands bandwidths rather than the subcarrier numbers play a critical role in PAPR distribution.
		Simulations were performed to illustrate that the proposed analytical expression of PAPR are applicable to the generic OFDM-based mixed-numerology system with a promising accuracy.
		It also demonstrates that PAPR is insensitive to the power fluctuation when the division of system bandwidth are determined.
	}
	
	{\color{black}
		Our future works will deal with the PAPR reduction techniques in mixed-numerology systems and the power allocation issue considering mean capacity maximization with PAPR constraint.
	}

	\section*{APPENDIX}
	
	By the derivation of (\ref{envelope}), the derivative of $ {r}(t) $ is
	\begin{equation}\label{}
	\dot{r}(t)=\frac{1}{r(t)}(x(t) \dot{x}(t)+\hat{x}(t) \hat{\dot{x}}(t)).
	\end{equation}
	Due to the linearity of the derivative operation, $ \dot{x}(t) $ and $ \hat{\dot{x}}(t) $ are also Gaussian random processes.
	From (\ref{moment1}) and (\ref{moment2}), the first and second order spectral moments of composite signal are denoted as 
	\begin{equation}\label{moment}
	\lambda_1 = \sum_{i=1}^{M}\eta_i{\alpha _i} 
	\quad \text{and} \quad
	\lambda_2 = \sum_{i=1}^{M}\eta_i{\beta _i},
	\end{equation}
	respectively.
	We denote $ {\mathbf{x}}=[{x}, \hat{x}, \dot{x}, \hat{\dot{x}}]^T $,
	where $ {x}, \hat{x}, \dot{x}, \hat{\dot{x}} $ are the variables of Gaussian process $ {x}(t),\hat{x}(t),\dot{x}(t),\hat{\dot{x}}(t) $.
	The joint PDF can be written as
	\begin{equation}\label{}
	f_{\mathbf{x}}(\mathbf{x})=\frac{1}{4 \pi^{2}\sqrt{|\mathbf{R}|}} \exp \left[-\frac{1}{2} \mathbf{x}^{T} \mathbf{R}^{-1} \mathbf{x}\right],
	\end{equation}
	where $ \mathbf{R} $ is the covariance matrix and can be written as
	\begin{equation}\label{}
	\mathbf{R}={1\over 2}\left[ \begin{array}{cccc}{1} & {0} & {0} & {\lambda_1} \\ {0} & \lambda_2 & {-\lambda_1} & {0} \\ {0} & {-\lambda_1} & {1} & {0} \\ {\lambda_1} & {0} & {0} & \lambda_2\end{array}\right].
	\end{equation}
	According to the probability theory, we have the conditional probability density function
	\begin{equation}\label{}
	\begin{split}
	&f_{\dot x\hat {\dot x}}\left( {\dot x,\hat {\dot x}|x = a,\hat x = b} \right) = \frac{f_{\mathbf{x}}(\mathbf{x})}{{f_{x\hat x}\left( {x = a,\hat x = b} \right)}} \\
	&= \frac{1}{{\pi \left( {{\lambda_2} - \lambda_1^2} \right)}}
	\exp \left[ { - \frac{{({\dot x}+b\lambda_1)^2 + ({\hat {\dot x}}-a\lambda_1)^2} }{{ {{{\lambda_2} - \lambda_1^2}} }}} \right],
	\end{split}
	\end{equation}
	which means
	\begin{equation}\label{}
	\begin{split}
	&\left( {\begin{array}{*{20}{c}}
		{\dot x}(t)\\
		\hat {\dot x}(t)
		\end{array}|x(t) = a,\hat {x}(t) = b} \right)  \\
	&\qquad\qquad\quad \sim {\mathcal N}\left( \lambda_1{\left[ {\begin{array}{*{20}{c}}
			{ - b}\\
			{a}
			\end{array}} \right],\frac{{{\lambda_2} - \lambda_1^2}}{2}\left[ {\begin{array}{*{20}{c}}
			1&0\\
			0&1
			\end{array}} \right]} \right).
	\end{split}
	\end{equation}

	Changing to polar coordinates with $ x = r\cos\phi $, $ \hat{x} = r\sin\phi $,
	we have
	\begin{equation}\label{}
	\begin{split}
	{\left(\cos \phi \dot{x}(t)+\sin \phi \hat{\dot{x}}(t) | x(t)=r \cos \phi, \hat{x}(t)=r \sin \phi\right)} \\ 
	\qquad\qquad\qquad\qquad\qquad\sim{ {\mathcal N}\left(0, \frac{1}{2}\left({{\lambda_2} - \lambda_1^2}\right)\right)}.
	\end{split}
	\end{equation}

	Then, we rewrite the expectation in (\ref{Rice}) as
	\begin{equation}\label{mean}
	\begin{split}
	&{E[|\dot{r}(t) \| r(t)=r]} \\ 
	&{=E\left[\frac{1}{r} | x(t) \dot{x}(t)+\hat{x}(t) \hat{\dot{x}}(t) \| x^{2}(t)+\hat{x}^{2}(t)=r^{2}\right]} \\ 
	&{=E_{\phi}[E[ | \cos \phi \dot{x}(t)+\sin \phi \hat{x}(t) \| x(t)=r \cos \phi}, \\ &\quad\quad
	{\hat{x}(t)=r \sin \phi ] ]}\\
	&= \sqrt{\frac{{{\lambda_2} - \lambda_1^2}}{\pi}}.
	\end{split}
	\end{equation}

	Using (\ref{enve_Ray}), (\ref{mean}), and (\ref{moment}) in (\ref{Rice}), we obtain 
	\begin{equation}\label{}
	\bar{U}(r, T_0) = T_0\sqrt{\frac{{{\lambda_2} - \lambda_1^2}}{\pi}} \cdot r  \exp \left({-r^2}\right),
	\end{equation}
	which finally leads to (\ref{equ1}).
	
	\bibliographystyle{IEEEtran}
	\bibliography{IEEEabrv,Ref_2019Spring}

\end{document}